\documentclass[graybox]{svmult}

\usepackage{mathptmx}       
\usepackage{helvet}         
\usepackage{courier}        
\usepackage{type1cm}        
\usepackage{makeidx}         
\usepackage{graphicx}        
\usepackage{multicol}        
\usepackage[bottom]{footmisc}

\makeindex             

\begin{document}

\title*{Tropical Cyclones as a Critical Phenomenon}
\author{\'Alvaro Corral}
\institute{\'Alvaro Corral \at 
Centre de Recerca Matem\`atica, 
Edifici Cc, Campus Bellaterra,
E-08193 Barcelona,
Spain,\\
\email{ACorral at crm dot cat}
}
%
%
\maketitle

\abstract{
It has been proposed that the
number of tropical cyclones as
a function of the energy they release
is a decreasing power-law function, 
up to a characteristic energy cutoff
determined by the spatial size of the
ocean basin in which the storm occurs.
This means that no characteristic scale
exists for the energy of tropical cyclones, 
except for the finite-size effects 
induced by the boundaries of the basins.
This has important implications for the 
physics of tropical cyclones.
We discuss up to what point tropical cyclones
are related to critical phenomena
(in the same way as earthquakes, rainfall, etc.),
providing a consistent picture of the energy balance
in the system.
Moreover, this perspective allows one to visualize
more clearly the effects of global warming on 
tropical-cyclone occurrence.
%
}


\section{Introduction}
\label{sec:1}

A fundamental way to characterize a physical phenomenon
is by analyzing the fluctuations of the 
energy it releases over successive occurrences.
Of course, in most of the cases this is not a simple issue.
For tropical cyclones, Bister and Emanuel \cite{Bister_Emanuel}
have found that the dissipated energy $E$ can be estimated
by integrating the cube of the surface velocity field over space and time, 
by means of the formula
\begin{equation}
E\simeq \int \rho C_D |v(\vec r,t)|^3 d^2 r \, dt,
\label{Bister}
\end{equation}
with $\rho$ the surface air density, 
$C_D$ the surface drag coefficient,
and $v(\vec r,t)$ the surface wind speed
at position $\vec r$ and time $t$.
It is implicit in the formula that the main contribution to 
dissipation comes from the atmospheric surface layer.

In order to obtain the distribution of energy then, 
one only needs to apply the previous formula
to as many tropical cyclones as possible
(without any selection bias)
and perform the corresponding statistics.
However, in practice, the available records do
not allow such a detailed calculation:
instead of providing a nearly instantaneous
velocity field, 
best-track data consist of a single value of the
speed reported every six hours
(the maximum sustained surface wind speed).

Emanuel has envisaged a way to reconcile the 
calculation of the energy with the limitation of the data
\cite{Emanuel_nature05}.
First, $\rho$ and $C_D$ can be approximated as constants
in Eq. (\ref{Bister}).
Second, one can apply
the similarity between radial profiles 
of speeds for different tropical cyclones to write 
$v(\vec r,t)=v_m(t)f(\vec r/R(t))$,
where $R(t)$ is the radius of storm at time $t$
(no matter how it is defined),
$v_{m}(t)$ is the maximum of the surface velocity field for all $\vec r$
at $t$,
and $f$ is the function that describes the shape
of the velocity profile (the same for all storms,
the scale given by $v_m$ and $R$).
This
yields a scaling between the integral over space on the one side
and the maximum speed and the radius on the other
(with the same constant of proportionality),
and then,
$$
E\propto \int  |v_{m}(t)|^3  R^2(t) dt,
$$ 
where
the symbol $\propto$ indicates proportionality.
An additional approximation is that the radius of the storm
is nearly uncorrelated with the speed, and therefore
assigning a common radius to all storms (al all times) leads only to random
errors in the evaluation of the energy.
Finally, enlarging the integration time step
up to $\Delta t =$ 6 hours gives
$$
E\propto PDI \equiv \sum_t v_t^3 \Delta t,
$$ 
with 
$PDI$ defining the so-called power dissipation index,
which is then a proxy for the total energy dissipated by a tropical 
cyclone during all its life.
The symbol $v_t$ denotes the maximum sustained surface wind speed.

A similar definition is that of the so-called accumulated cyclone energy ($ACE$)
\cite{Bell,Gray_HDP}, which integrates kinetic energy over time,
$$
ACE \equiv \sum_t v_t^2 \Delta t,
$$
where the essential difference with the $PDI$ is the replacement 
of the cube of the speed by a square.
Note that the time integral of the kinetic energy
is not an energy, unless some proportionality factors are introduced
in the formula, in the same way as in Eq. (\ref{Bister}).
In any case, in this work
we will study the distribution of energy dissipated by tropical cyclones using 
both $PDI$ and $ACE$ as proxies for the energy,
evaluated over the complete lifetime of the storms.


\section{Power-Law Distribution of the Energy of Tropical Cyclones}
\label{sec:2}

In order to describe probability distributions we will use
the probability density function.
For the case of power dissipation index this is defined as 
the probability that the value of this variable lies in a narrow interval 
of size $dPDI$ around a concrete $PDI$,
divided by $dPDI$ to make the result independent on $dPDI$,
i.e., 
$$
D(PDI) \equiv \frac{ \mbox{Prob[ $PDI-dPDI/2 \le $ value  $ < PDI +dPDI/2$ ]}}{dPDI},
$$ 
where Prob denotes probability,
which is evaluated as the number of events that fulfill the condition
divided by the total number of events.
This definition ensures normalization, 
$\int_0^\infty D(PDI) dPDI =1$.
Note also that the units of the density are the reciprocal of the units
of the variable, so, for $D(PDI)$ these are s$^2/$m$^3$
(if the $PDI$ is measured in m$^3$/s$^2$).
An analogous definition applies to the probability density of the $ACE$,
$D(ACE)$, or of any other variable.

Recently, we have shown that the distribution of $PDI$
in different tropical-cyclone basins
follows a power law, 
$$
D(PDI) \propto 1/PDI^\alpha,
$$
except for the largest and smallest values of $PDI$.
The exponent $\alpha$ turns out to be close to 1
(between 1 and 1.2, roughly) \cite{Corral_hurricanes}.
Note that an exponent equal to one implies that 
all decades contribute in the same proportion
to the total number of events, 
in other words,  
any interval of $PDI$ values in which the extremes
keep the same proportion contains the same probability.

The calculation of $D(PDI)$ is not direct, though.
The quantity of interest, the $PDI$, varies across
a broad range in the basins studied, from less than $10^9$ m$^3$/s$^2$
to more that $10^{11}$ m$^3$/s$^2$, being necessary to 
plot the distribution in logarithmic axes in order to 
represent the different scales.
Moreover, this has the advantage that 
on a log-log plot 
a power law appears as a straight line
(note that this is not the case for the cumulative distribution
function if the power law has an upper cutoff
\cite{Hergarten_book}).

On the other hand, the broad range of variation also makes inappropriate
the use of a constant interval size $dPDI$
(essentially, $dPDI$ should be large enough to contain enough
statistics but small enough to provide a complete sampling of the range
of variation of $D(PDI)$).
Logarithmic binning is a solution to this problem
\cite{Hergarten_book}, where the size of the bins appears
as constant in the logarithmic scale used.
(An equivalent, simpler solution, is to work with the distribution 
of $\ell \equiv \ln PDI$,
calculating its probability density $D(\ell)$ 
using standard linear binning
and then obtaining the 
$PDI$ density by means of the change of variable $D(PDI)=D(\ell)d\ell/dPDI=D(\ell)/PDI$;
of course, $D(PDI)$ and $D(\ell)$ have different functional forms, 
despite the ambiguous notation.)
Naturally, similar considerations hold for the distribution of $ACE$.

Turning back to the results of Ref. \cite{Corral_hurricanes}, 
it is shown there that the $PDI$ distribution is 
well described by a decreasing power law
in the North Atlantic (NAtl), 
the Northeastern and Northwestern Pacific (EPac and WPac), 
and the Southern Hemisphere (SHem) basins,
with an exponent $\alpha$ ranging from $0.98 \pm 0.03$ in the WPac
to $1.19\pm 0.07$ in the NAtl,
where the uncertainty refers to one standard deviation
of the maximum-likelihood-estimator mean value.
The power law holds from a range of a bit more than one decade
(for the NAtl and the EPac) up to two decades (for the WPac).
The data used were the best tracks from NOAA's National Hurricane Center
for NAtl and EPac  \cite{NOAA,NOAA_hurdat}
and from US Navy's Joint Typhoon Warning Center for WPac and SHem
\cite{ATCR_report,JTWC_data}.
Here we will use the same data sets.

\begin{figure}[b]
\sidecaption
\includegraphics[scale=.55]{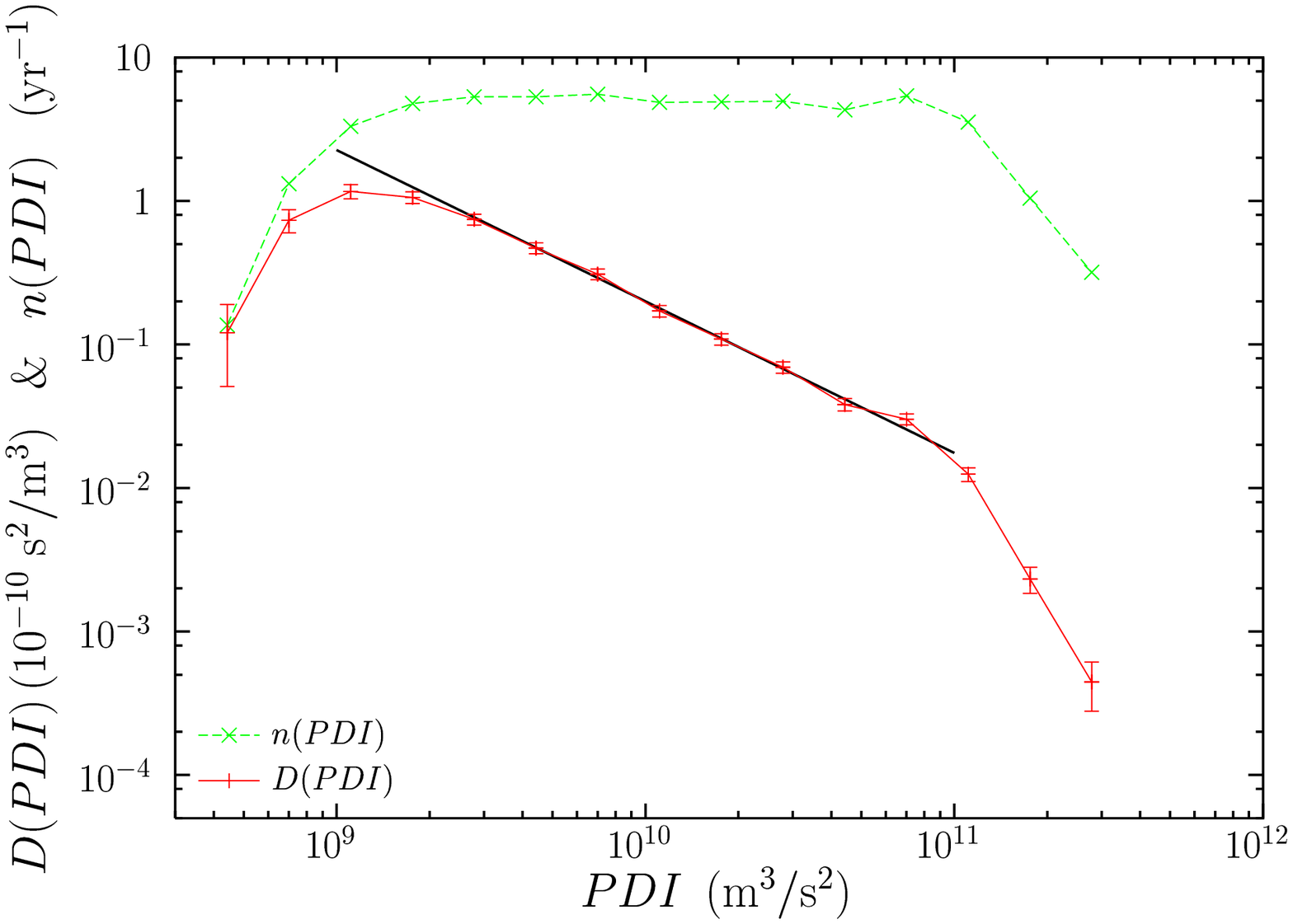}
\includegraphics[scale=.55]{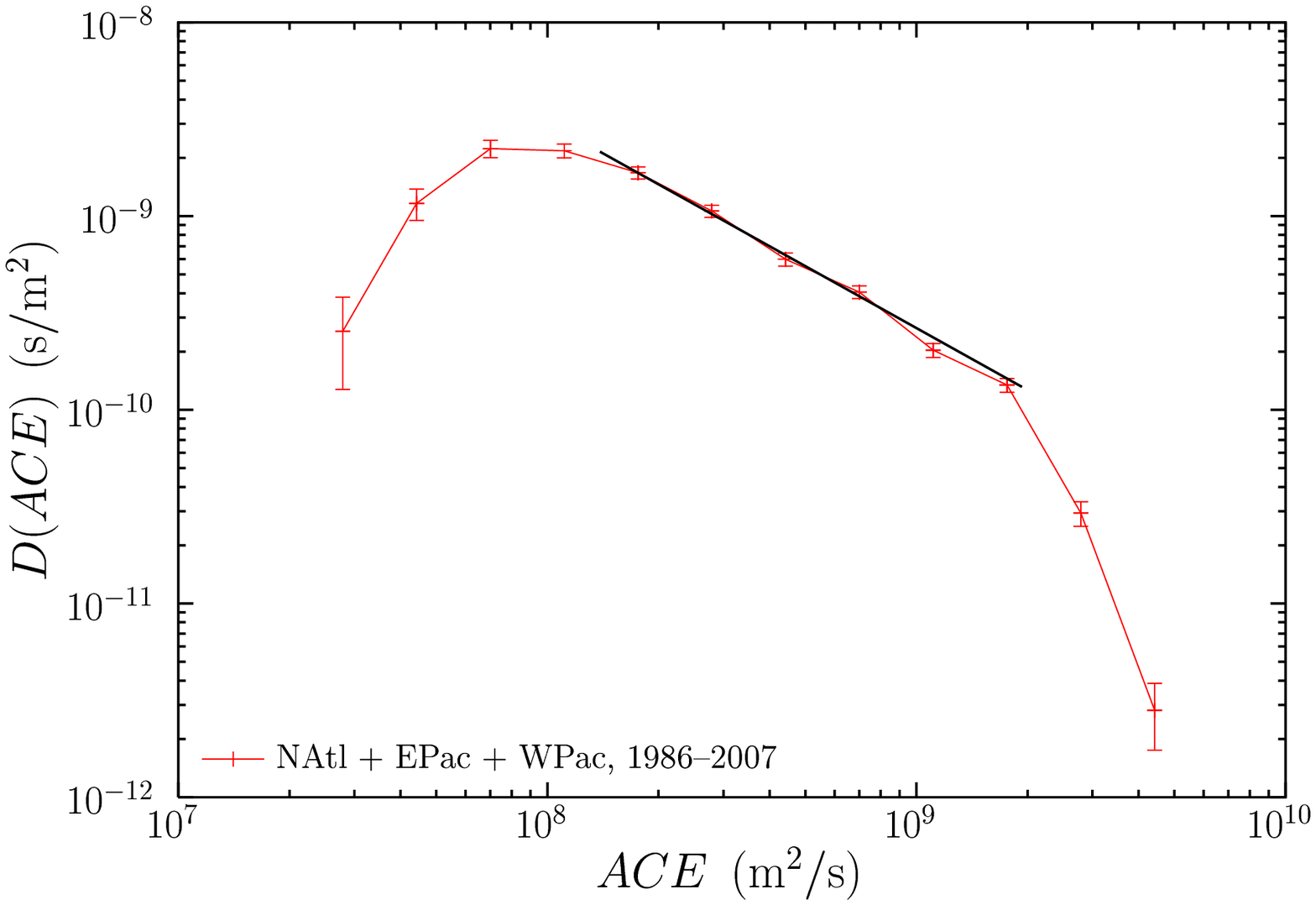}
\caption{
Probability density of tropical-cyclone dissipated energy,
for the NAtl, EPac, and WPac basins, during 1986--2007,
compressing 1212 storms.
(a) $PDI$ probability density, $D(PDI)$,
together with the number of tropical cyclones
in each bin per year $n(PDI)$.
A maximum-likelihood fit of the distribution 
between the values $3.2 \cdot 10^9$
and $4.0 \cdot 10^{10}$ m$^3/$s$^2$
yields $\alpha=1.07 \pm 0.06$
with a $p-$value $= 99   \pm 0.3  \%$
(calculated generalizing the method of Ref. \cite{Clauset},
with a resolution in the optimization of 10 points per decade). 
(b) $ACE$ probability density.
A maximum-likelihood fit 
between the values $1.6 \cdot 10^8$
and $2.0 \cdot 10^{9}$ m$^2/$s.
yields $\alpha=1.10 \pm 0.05$
with a $p-$value $= 59 \pm 1.6 \%$.
}
\label{fig:1}       
\end{figure}

As an illustration, we show in Fig. \ref{fig:1}(a)
the $PDI$ distribution for the Northern Hemisphere
(excluding the Indian Ocean, i.e., only NAtl+EPac+WPac)
for the years 1986 to 2007.
Tropical depressions (storms whose maximum sustained surface wind speed
does not exceed 34 knots), not included in the NAtl and EPac records,
have been eliminated from our analysis of the WPac,
for consistency. 
Of course, the results are in agreement with Ref. \cite{Corral_hurricanes},
with an exponent $\alpha=1.07 \pm 0.06$.

\begin{figure}[b]
\sidecaption
\includegraphics[scale=.55]{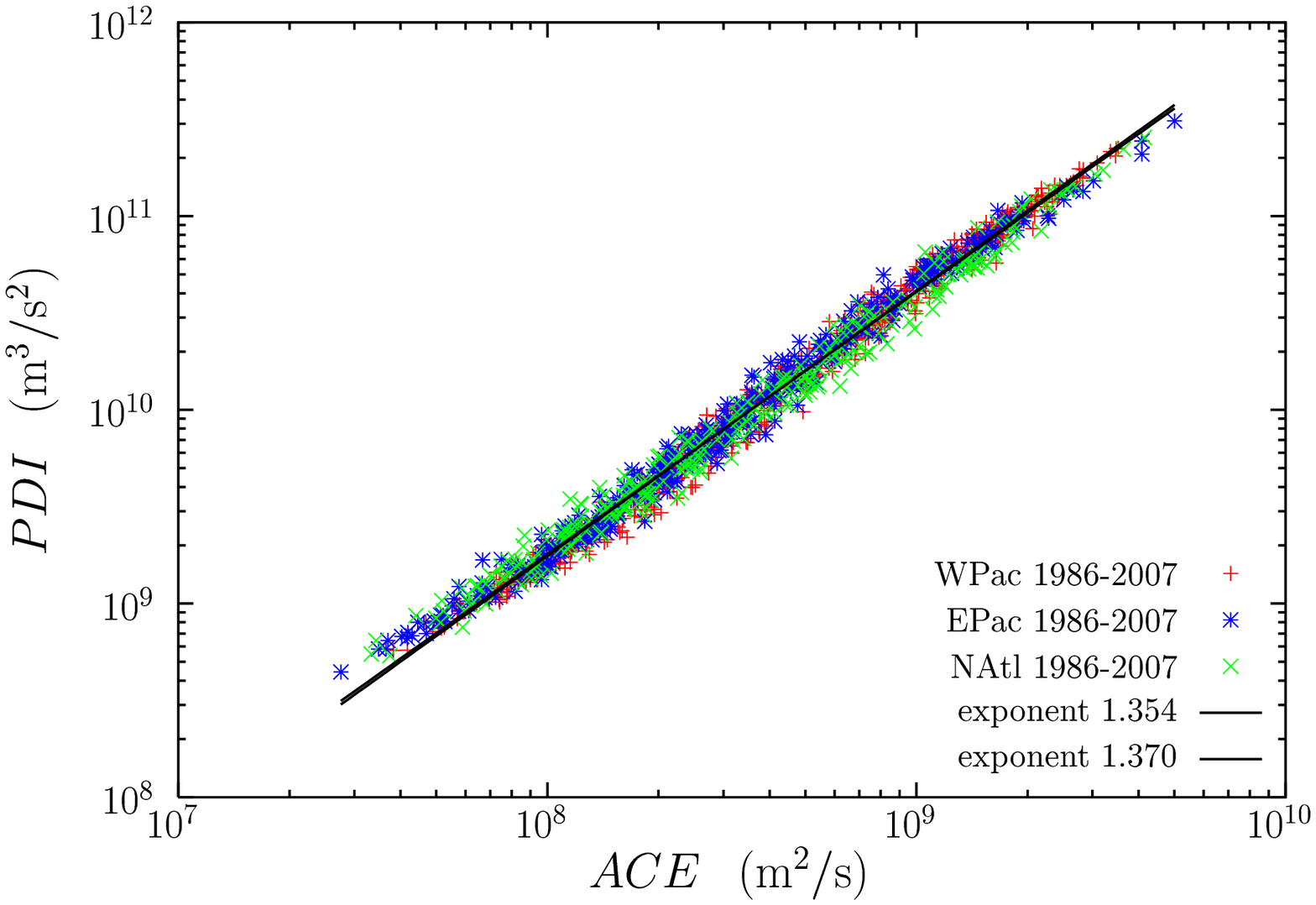}
\caption{
Non-linear correlation between $PDI$ and $ACE$.
The former variable is plotted versus the latter 
for all hurricanes, typhoons, and tropical storms
occurring in the NAtl, EPac, and WPac for the period
1986--2007.
A power-law correlation $PDI \propto ACE^{1.36}$
shows up, with a corresponding linear correlation 
coefficient (for the logarithm of the variables)
$\rho=0.994$.
The two regression lines are included,
although they overlap.
}
\label{fig:scatter}       
\end{figure}

If instead of the $PDI$ we use the $ACE$ the 
results do not change in essence, 
yielding $\alpha=1.10\pm 0.05$,
as displayed in Fig.  \ref{fig:1}(b).
The reason for the coincidence of results
between both variables
is due to the fact that they
are highly (though non-linearly) correlated.
Figure \ref{fig:scatter} shows a scatter plot for the values
of the $PDI$ versus the $ACE$,
using the same data as above.
A linear regresion applied to $\ln PDI$ versus $\ln ACE$
yields $PDI \propto ACE^\gamma$, with $\gamma \simeq 1.36$ 
and a correlation coefficient $\varrho=0.994$.
If we write our probability distributions as
$D(PDI)\propto 1/PDI^{1+\beta_p}$ and
$D(ACE)\propto 1/ACE^{1+\beta_a}$,
and introduce $PDI \propto ACE^\gamma$
in the identity relation $D(PDI) = D(ACE) dACE/dPDI $,
one gets $\beta_a = \gamma\beta_p$, 
which implies that a power law with exponent $\alpha=1$
(i.e., $\beta=0$)
is invariant under power-law changes of variables.
These results are in good concordance with our 
numerical findings.

\begin{figure}[b]   
\sidecaption
\includegraphics[scale=.65]{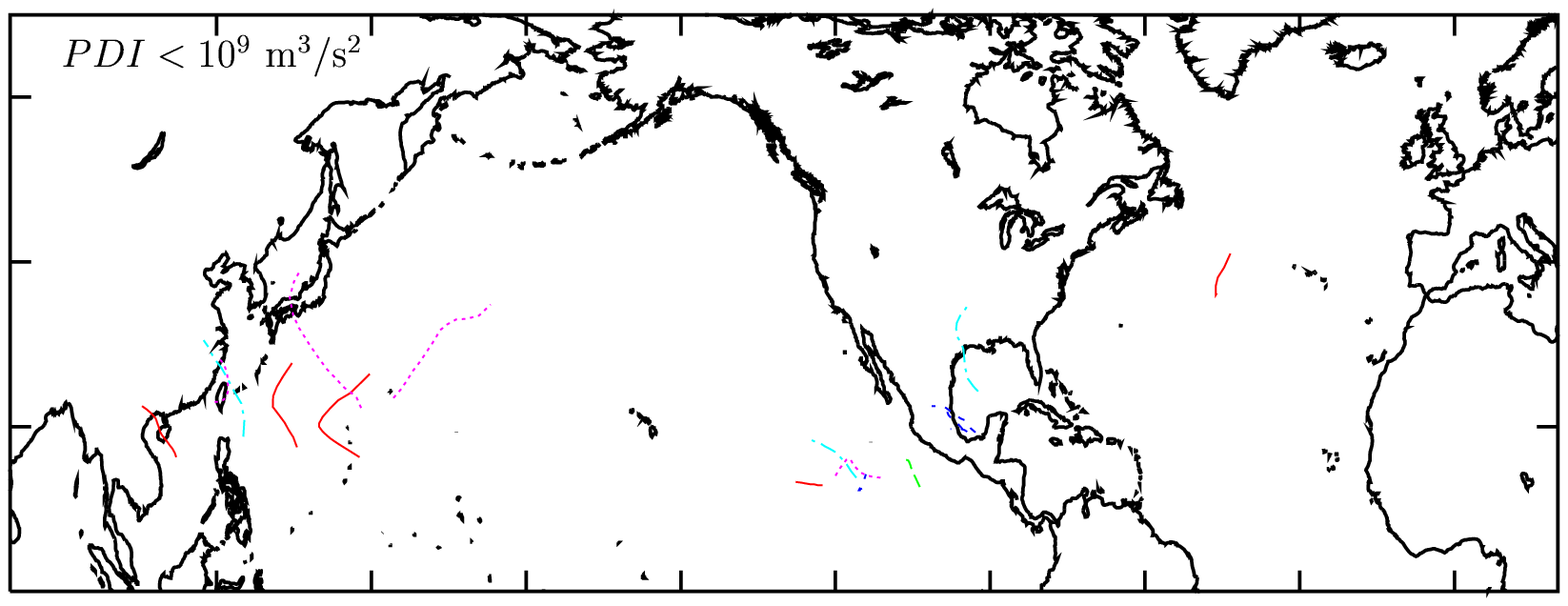}\\
\includegraphics[scale=.65]{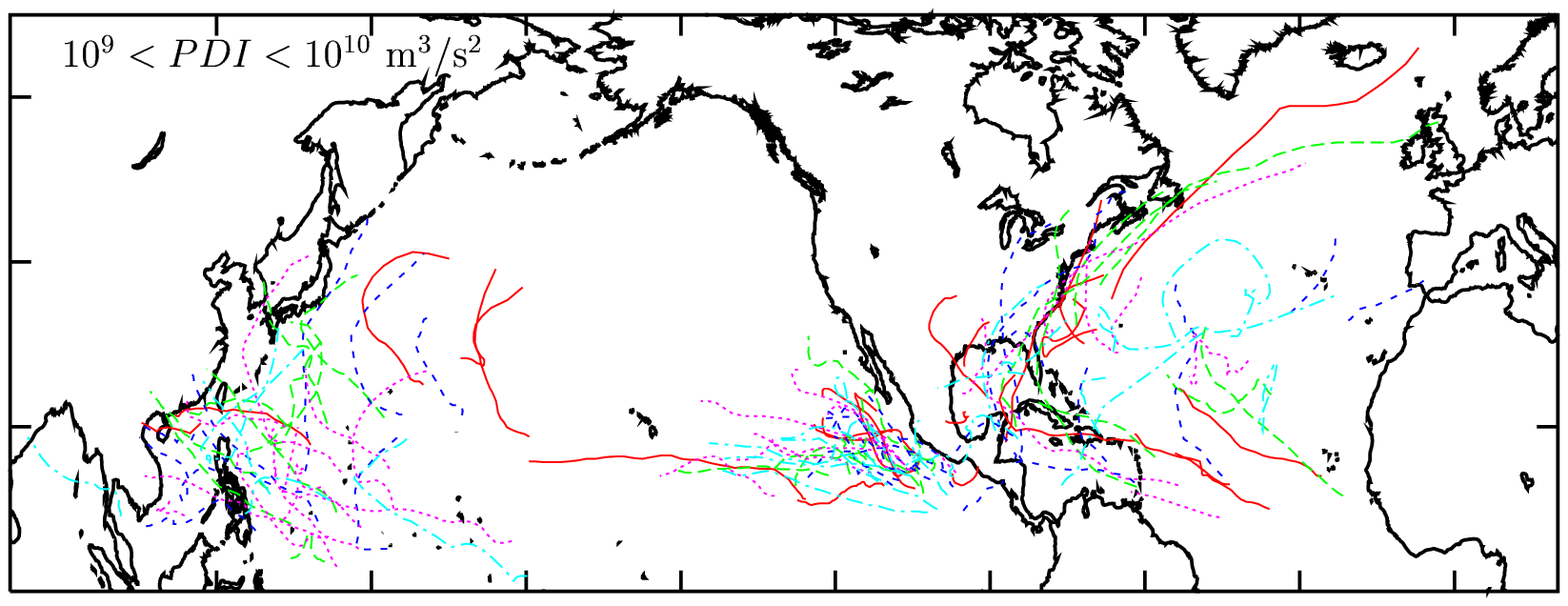}\\
\includegraphics[scale=.65]{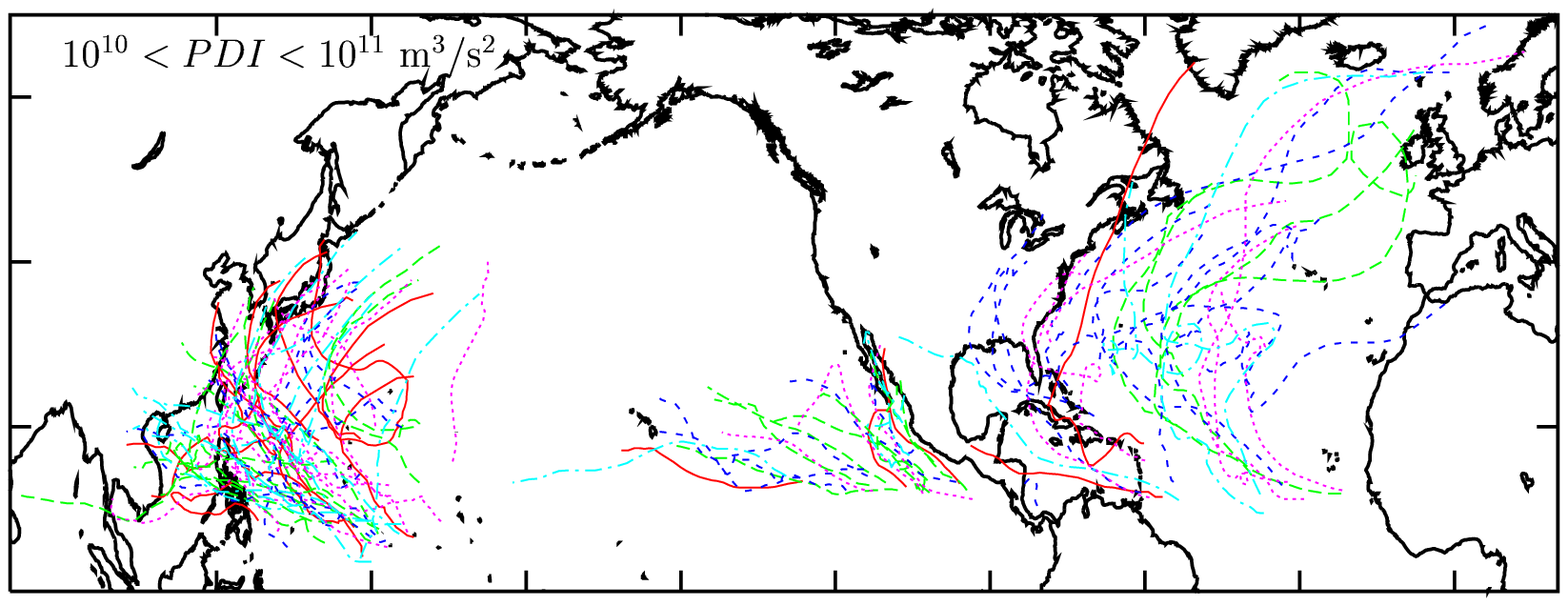}\\
\includegraphics[scale=.65]{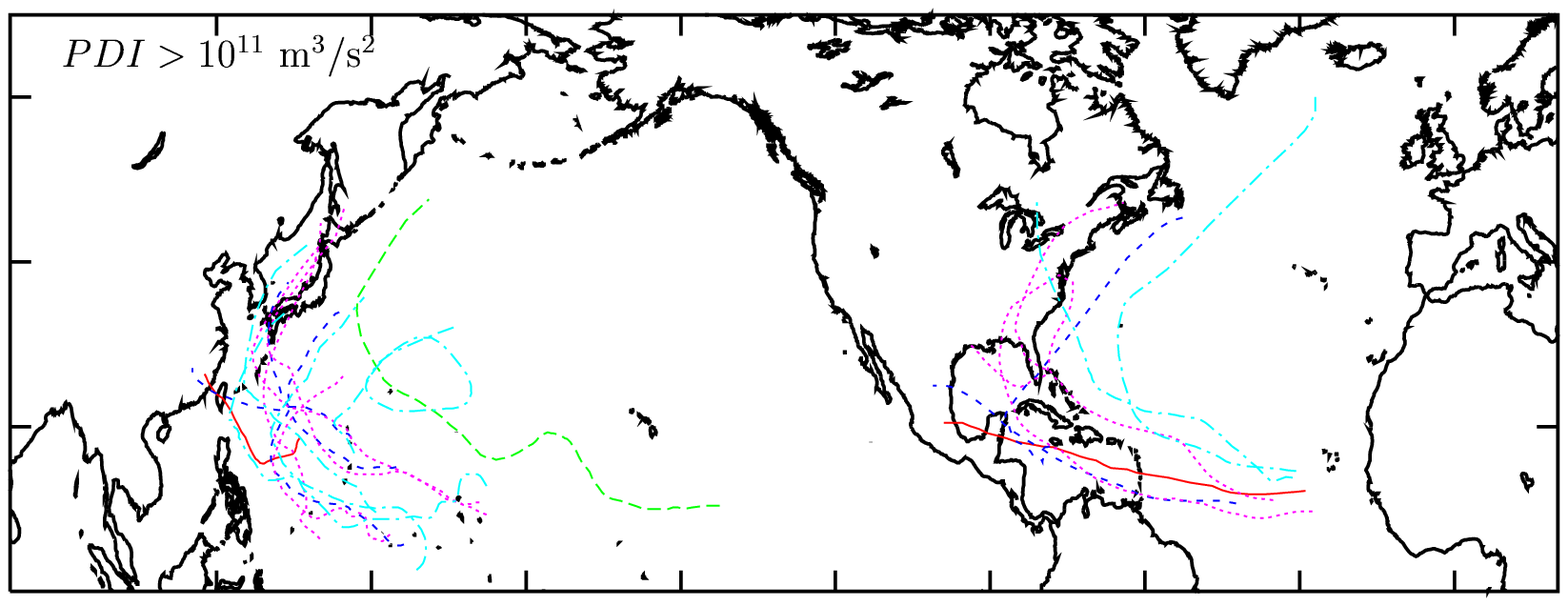}\\
\caption{
Trajectories of hurricanes, typhoons, and tropical storms (i.e., excluding tropical 
depressions) in the WPac, EPac, and NAtl, separated for different values of the $PDI$.
Different line styles correspond to different years, from 2003 to 2007.
The top panel points to the incompleteness of the record for the smallest $PDI$ values, 
whereas the bottom panel shows that the tropical cyclones with the largest $PDI$ values
tend to travel through the whole basin.}
\label{fig:2}       
\end{figure}

An important issue is the deviation of the distribution from the power-law behavior
for small and large energies.
In the first case, the deviation is due to the scarcity of data referring
to small storms. The best tracks of the National Hurricane Center
(for NAtl and EPac) do not contain tropical depressions, so the data
are truncated including only hurricanes (of category 1 at least)
and tropical storms (maximum sustained surface wind speed larger 
than 34 knots but below hurricane category).
In the case of the best tracks of the Joint Typhoon Warning Center
some tropical depressions are included, 
but these are very few, only those ones which are consider ``significant''.
This artificial truncation of the data obviously makes the distribution
depart from the power-law behavior for small values of the energy.
The paths of the tropical cyclones of a small part of the record,
just for the 5 year period 2003-2007,
are shown in Fig. \ref{fig:2},
separated in distinct $PDI$ ranges;
the top panel shows how the set of storms with $PDI < 10^9$ m$^3/$s$^2$
seems certainly incomplete.

More fundamental is a decay faster than power law for large values of the energy.
We suggest in Ref. \cite{Corral_hurricanes} that this is due to a finite size effect:
the spatial size of the basin is not big enough to sustain tropical cyclones
with larger $PDI$ values
(remember that the $PDI$ integrates $v_t^3$ over time, 
and time can be considered equivalent to space). 
Indeed, when a tropical cyclone in any of the basins considered reaches a
$PDI$ of about $10^{11}$ m$^3$/$s^2$ it is very likely that its evolution
is affected by the boundaries of the basin, which are constituted essentially
either by continental land or by 
a colder environment;
this deprives the tropical cyclone from its source of energy
in the form of warm water
and provokes its attenuation and eventual death, see Fig. \ref{fig:2}(d). 
Under the name ``colder environment'' 
we can compress both a low sea surface temperature
(as it happens in high latitudes or
with the cold California Current)
or the presence of extratropical weather systems.
In any case, this makes the boundaries of the basin be not fully ``rigid''
in contrast to condensed matter physics
(even for the case of continental boundaries, 
there have been hurricanes that have jumped 
from the NAtl to the EPac).



Another relevant issue to take into account
is up to what point the power law is the right distribution
to fit the $PDI$ and $ACE$ distributions.
We have shown in Ref. \cite{Corral_hurricanes} that the power law
provides indeed a good fit, but this does not exclude that
other distributions can fit the data equally well, 
or even better.
In fact, any power law in a finite domain
(let us say, with the variable $x$ in between $m_1$ and $m_2$) 
can be fit also by a lognormal distribution, 
with a $\sigma$-parameter (the standard deviation
of the underlying normal distribution) which tends
to infinity.
Indeed, the lognormal probability density, 
$$
D(x) \propto \frac 1 x \exp\left( -\frac {(\ln x -\mu)^2}{2 \sigma^2}\right),
$$
can be written as
$$
D(x) \propto \frac 1 {e^\mu} \left(\frac {e^\mu} x \right)^{1+(\ln x - \mu)/(2\sigma^2)}.
$$
This is a kind of pseudo-power-law,
with a pseudo-exponent 
$1+(\ln x - \mu)/\sigma^2$ 
that changes very slowly with $x$ if $\sigma$ is big enough.
Taking 
$m_1\equiv e^{\mu+\sigma^2 \epsilon_1}$ 
and $m_2\equiv e^{\mu+\sigma^2 \epsilon_2}$,
the pseudo-exponent changes from $1+\epsilon_1$ for $x=m_1$
to $1+\epsilon_2$ for $x=m_2$, 
and the extreme values $m_1$ and $m_2$ can be as large as desired
if $\sigma$ is big enough%
\footnote{The last two sentences have been corrected, 
in comparison with the printed version of the chapter.}.
Note, anyhow, that the lognormal distribution has two parameters
($\mu$ and $\sigma$),
whereas the power law has just one ($\alpha$).
A similar argument can be presented for other long-tailed distributions.
In conclusion, the choice of the right fit is a problem that cannot be
solved only by means of the statistical analysis, and 
it is the physical knowledge which has to provide 
a criterion to select the most appropriate distribution.
The rest of this work will justify the preferability of the power law
as a physical model of tropical-cyclone dissipation distribution.

\section{Power-Law Distribution of Earthquake Energies}

The power-law distribution of energy dissipation in tropical cyclones is analogous
to the well known Gutenberg-Richter law of earthquakes.
Let us see how.
This law states that, for a given spatial region and over a certain period of time, 
the number of earthquakes with magnitude 
larger than $M$ is about 10 times greater than the number of earthquakes
with magnitude larger than $M+1$, 
which in its turn is 10 times greater than the number of earthquakes
larger than $M+2$ and so on \cite{Kanamori_rpp}.
In mathematical terms, the number of earthquakes above $M$,
denoted by $N(M)$, is a decreasing exponential function,
$$
N(M) \propto 10^{-b M}
$$
where the $b-$value is a parameter close to 1.

The cumulative distribution function of magnitudes,
defined as $S(M) \equiv $ Prob[  magnitude value  $ \ge M $ ]
is estimated directly from $N(M)$ as
$S(M) = N(M)/ \mathcal{N}$,
where $ \mathcal{N}$ is the total number of earthquakes considered, 
of any magnitude.
It is obvious that the cumulative distribution is exponential, 
and therefore the density, given as $D(M)= - dS(M)/dM$,
is an exponential too, with the same $b-$value.
(This allows that, when working with magnitude distributions,
one does not need to specify if one is measuring the density or the 
cumulative distribution, unfortunately.
Of course, this is only acceptable for exponential distributions).

But magnitude is not a physical variable (it has no units).
It is believed, at least as a first approximation, that the energy $E$ radiated 
in an earthquake is an exponential function of the magnitude, 
$E \propto 10^{3M/2}$ 
(with a proportionality factor between $10^3$ and $10^5$ Joules)
\cite{Kanamori_rpp}.
Therefore, the energy probability density will be a power law,
$$
D(E) = D(M)dM/dE \propto D(M)/E \propto 1/E^{1+2b/3}.
$$
Note that, as in the case of tropical cyclones, 
we have used the same symbol for the density of energies
and for the density of the logarithm, although the functional
form of each one is not the same
(power law versus exponential, respectively).

Summarizing, although the Gutenberg-Richter law implies an exponential
distribution of the magnitudes of earthquakes, in terms of the radiated energy
the Gutenberg-Richter law is given by a power law. 
Then, the fundamental difference in the 
structure of energy release between earthquakes and tropical cyclones
is only quantitative and not qualitative,
as both phenomena follow power-law distributions of energy 
%
with $\alpha \simeq 1.7$ in the first case and $\alpha\simeq 1.1$ for tropical cyclones.
Another difference is the deviation from the power-law behavior
at the largest values of the energy in tropical cyclones;
in the case of earthquakes
the existence or not of this boundary effect is not clear 
\cite{Kagan_calcutta,Main_ng}.

In fact,
many other complex phenomena in the geosciences yield power-law distributions
of energies, or, broadly speaking, ``sizes''.
These phenomena include, in addition to earthquakes and tropical cyclones:
rainfall \cite{Peters_prl},
landslides and rock avalanches \cite{Malamud_hazards,Frette96},
forest fires \cite{Malamud_science},
volcanic eruptions \cite{Lahaie}, 
solar flares \cite{Arcangelis,Baiesi_flares},
the activity of the magnetosphere \cite{Wanliss},
tsunamis \cite{Burroughs}, 
and perhaps meteorite impacts \cite{Chapman}.
Nevertheless, the power laws are not totally ubiquitous,
see for instance Ref. \cite{Corral_fires}.

\section{Relevance and Mechanisms for Power-Law Distributions}

Which are the implications of having a power-law distribution,
as it happens for the released energy of tropical cyclones, earthquakes
and other phenomena just mentioned?
In general, power-law distributions denote the presence of three main characteristics:
\begin{itemize}
\item Divergence of the mean value of the variable.
\item Absence of a characteristic scale.
\item Possible connection with criticality.
\end{itemize}
Let us explain each one.

\subsection{Divergence of the mean value}

Regarding the first issue, it is obvious that, if we consider
the mean energy value, this fulfills $\langle E\rangle \equiv 
\int_m^\infty E D(E) dE = \infty$, if the power law exponent
$\alpha$ is smaller than 2 (but larger than 1 for normalization),
with $m$ the minimum value of the energy.
Note that this is a property which is neither 
a characteristic of all power laws
nor
exclusive of some power laws
(there are many other distributions which show this divergence,
for instance, $D(E) \propto E^{-\alpha} \cos^2 E $).

Obviously, from a physical point of view, the mean energy dissipated 
by a phenomenon as earthquakes or hurricanes cannot be infinite (the Earth has a finite
energy content) and therefore the power-law behavior cannot be extrapolated
to infinity. But if we do not know up to which maximum energy value the power
law holds (which seems to be the case of earthquakes, but not that of tropical cyclones),
the mean value of the energy is not defined and its calculation from 
any data set does not converge.
What happens is that the scarce extreme events dominate 
the calculation of the average, 
as when they occur their contribution to the mean value 
is large enough to alter significantly this mean value.
So, the fluctuations are 
the most significant trend of the energy,
and not the mean value. 
In the case of tropical cyclones, we can only say that
if it were not for the finite-size effects imposed 
by the boundaries of the basins, the mean released
energy could not be calculated.

\subsection{Lack of characteristic scale}
 
\begin{figure}[b]
\sidecaption
\includegraphics[scale=.55]{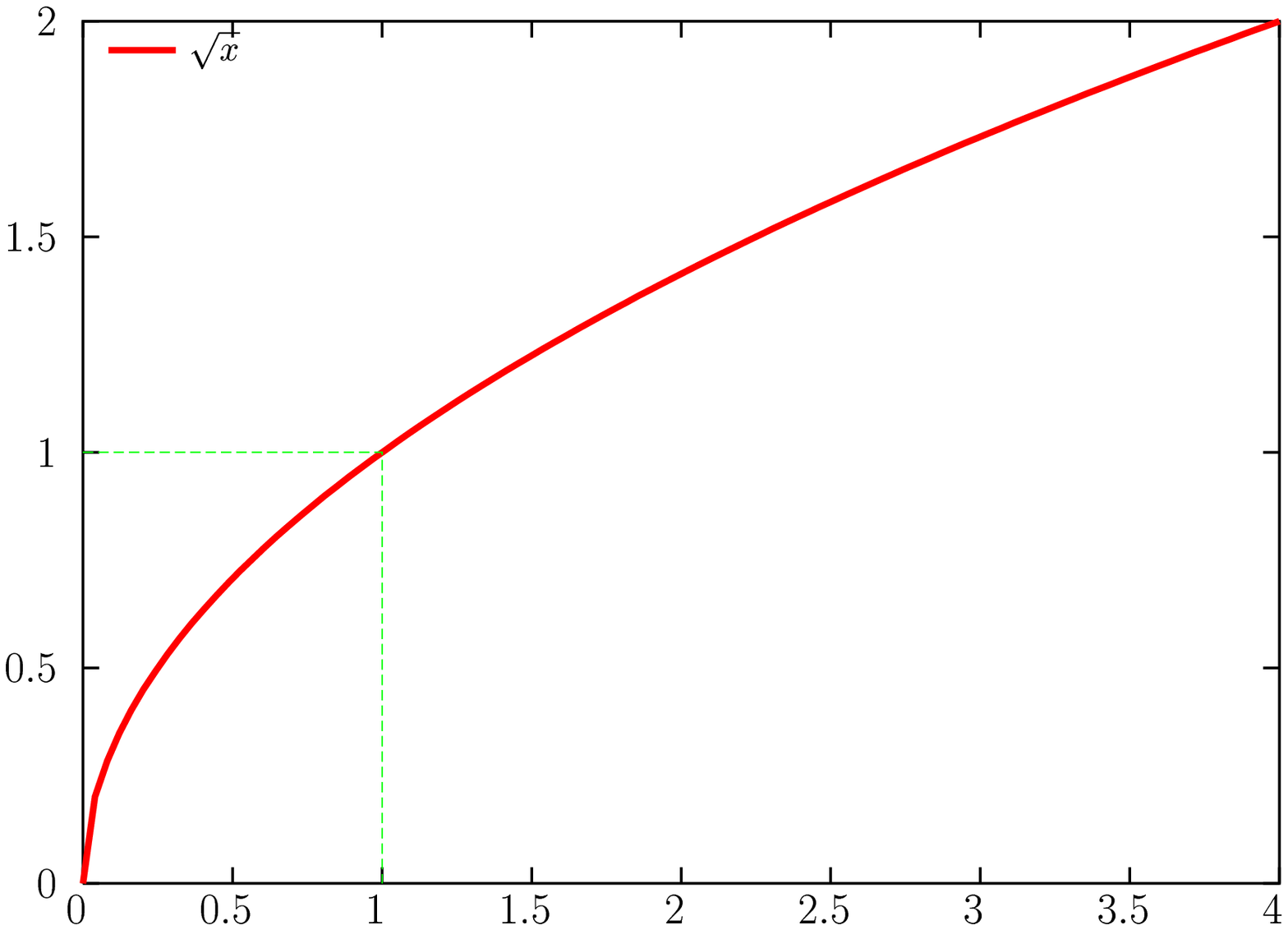}
\includegraphics[scale=.55]{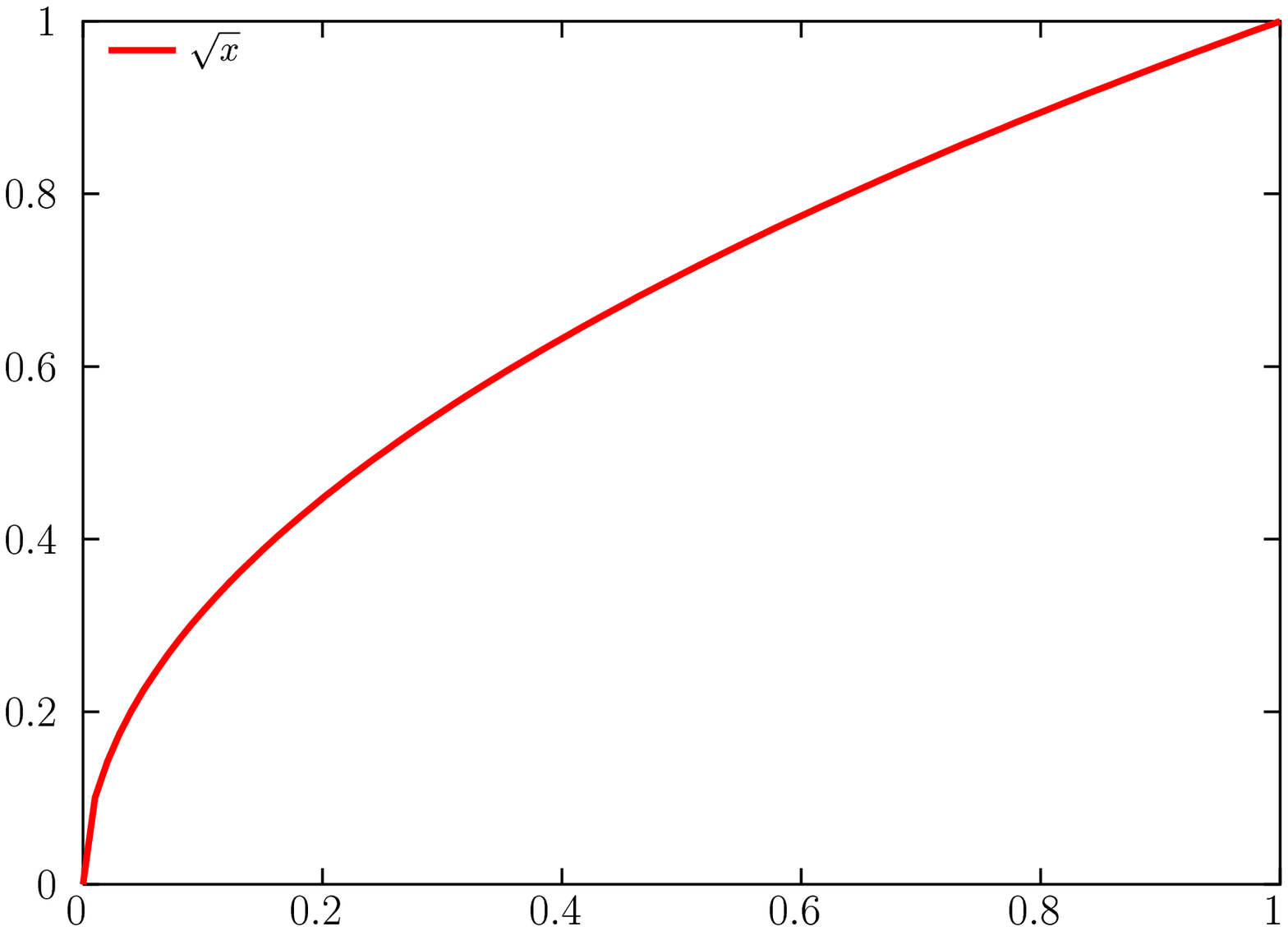}
\caption{
Illustration of the scale invariance of power laws.
A power-law function with $\alpha=-0.5$ looks exactly the same at two different scales.
The box in the top panel is enlarged in the bottom one,
the scale factors of the scale transformation are $a=4$ and
$c=1/a^\alpha =2$.
}
\label{fig:3}       
\end{figure}

In contrast to the first one,
the second property, the absence of any characteristic scale for the energy release,
is an exclusive property of power laws \cite{Christensen_Moloney}.
It is possible to show that a power-law function $g(x)\propto 1/x^{\alpha}$
(with $-\infty < \alpha < \infty$)
is the only solution to the scale-invariance condition:
$g(x)=c g(x/a), \, \forall x, \forall a$,
where it turns out then that $c$ has to be related to $a$ by $c=1/a^\alpha$
(alternatively, fixing the relation between $a$ and $c$ determines the value of the 
exponent $\alpha$).
This condition means that it does not matter in which scale we
look at the variable $x$, we will see the same shape for the function $g(x)$.
For example, let us take $a=1000$, then, when we write $x/a$
we are looking at $x$ at a scale that is 3 orders of magnitude smaller 
than the original one (we go from the scale of kilo-Joules to Joules, let us say);
if we take $c=0.01$, which means that we are performing another linear 
transformation in the $y-$axis, we find that the corresponding scale-invariant function
is $g(x)\propto 1/ \sqrt[3]{x^2}$, indeed, $\alpha=-\log c/\log a = 2/3$.
Another example is given in Fig. \ref{fig:3}.

Nevertheless, there is a ``little'' problem regarding scale invariance of 
probability distributions: a function of the kind $g(x) \propto 1/x^\alpha$
cannot be a probability density for all $x$, even for just $x>0$, as 
$\int_0^\infty g(x) dx = \infty$, for all $\alpha$.
In practice, it is necessary a small-energy cutoff $m$ if $\alpha \ge 1$
or a large-energy cutoff if $\alpha \le 1$, so, the scale invariance only can exist
for a range of $x$ and $a$, and not for all of them.

\subsection{Criticality}

One has to recognize that scale invariance is a rather strange property.
How can it be that the relative proportion of the value of a physical observable
at two different values of its variable,
$g(x)/g(x/a)$,
is the same ($=c$)
at the milli-scale and at the Mega-scale
(i.e., independent on $x$)
if the values $x$ and $x/a$ take a constant proportion ($a$)?
This means that the study of the function $g$ does not allow us to distinguish
the scale of observation
(in order words, $g$ cannot be used as an $x-$meter).
It seems obvious that the same physics has to operate at very different scales.


In order to 
elucidate how scale invariance in energy distributions is achieved
we need a model of the energy release.
Let us have in mind the case of earthquakes, just to fix ideas.
There, energy is released in tectonic faults in an avalanche-like process.
The picture can be summarized as follows:
stress in the Earth crust displays very small changes;
sooner or later,
at some point in the crust, 
the static friction cannot sustain a small variation in stress
and a slip takes place; this local slip increases the stress in the neighboring area, 
where more slip can be induced in this way, and so on.

A very simple model of this process is given by a chain of dominoes:
the slip at a fault patch is represented by the toppling of one piece;
a sequence of topplings, until the end of the activity,
constitutes an avalanche that represents an earthquake
(or other phenomenon); and
the energy released in the process will be proportional 
to the number of topplings, which is called the avalanche size.
In the usual game, the toppling of one piece induces the toppling of the next 
and so on; this is the so-called {\it domino effect} and yields toppling events (avalanches) 
whose size is equal to the size of the system
(the number of topplings equals the total number pieces).
We arrive then to the so-called characteristic-earthquake model.
But this is not what the data tells us;
the Gutenberg-Richter law shows that there should be avalanches of all sizes, 
with no characteristic scale.

We need to modify the domino model.
Instead of having that one toppling always induces just one toppling, 
let us consider that one toppling induces one toppling, or none, 
or may be two, or three, etc.
That is, we have a random number of topplings, 
with the probability of the number of induced topplings
given by the same probability distribution for all pieces.
For this purpose it is convenient to imagine not a one-dimensional domino 
chain but an array of pieces.
Mathematically, this is just an image of a simple {\it branching process},
introduced in science to describe in the first place the growth and extinction
of populations.

The outcome of a branching process depends on the so-called branching ratio $B$, 
which is the average number of topplings induced directly
by a single toppling (from one time step to the next)
\cite{Harris,Sornette_critical_book}.
It is clear that if $B>1$ the process will have a tendency to grow exponentially,
giving rise to a system-spanning avalanche
(although there is also a finite probability that the chain of topplings 
dies spontaneously);
in contrast, if $B<1$, the activity attenuates fast, on average,
and the size of the avalanches is small.
However, if $B$ is precisely equal to one (with a standard deviation different than zero) 
these two tendencies compensate, 
and then we loose any characteristic scale in the size of the avalanches:
they are power-law distributed, the concrete shape of the density being
$D(s) \propto 1/s^{3/2}$, with $s$ the size of the avalanche (proportional to the energy).
This case is called a {\it critical branching process}.

So, in principle, we have arrived at a reasonable model to generate power-law distributions,
just adjusting the branching ratio to be equal to one.
The next question in order to give an explanation of these phenomena
is of course how the fine tuning of the branching ratio is achieved in nature.
An answer is given by the idea of {\it self-organized criticality} proposed
by Bak {\it et al.} in the 80's
\cite{Bak_book,Jensen}.
The basic idea is the existence of a feedback mechanism that keeps
the branching ratio close to one;
if it is larger than one this produces large avalanches 
and in this case the branching ratio is reduced
(the distance between the domino pieces is increased somehow);
if the branching ratio is small, the avalanches are small, 
and then the branching ratio is increased
\cite{Zapperi_branching}.

The idea is better illustrated substituting the toppling of 
domino pieces by the toppling of grains in a sandpile
\cite{BTW87,Christensen_oslo}.
The advantage of the sandpile is that after a large avalanche
(which usually happens for high $B$),
many grains leave the pile through its open boundaries
(the pile is built over a finite support) and
this decreases the average slope of the pile, 
making more difficult the toppling of the remaining grains and
reducing then the branching ratio.
On the contrary, when small avalanches predominate
(low $B$) the grains do not reach the boundaries
of the pile, and a subsequent slow addition of more
grains increases the slope and also
the branching ratio (as the toppling of the grains is 
facilitated by a steeper pile).

The ideal sandpile is a particular realization
of one kind of systems called {\it slowly driven,
interaction-dominated threshold systems} \cite{Jensen},
whose three main ingredients are, as the name denotes:
a slow energy input, an intermediate energy storage
caused by local thresholds, and
sudden bursty energy releases that 
spread through the system
\cite{Peters_JH}.
The energy input comes from the slow addition of grains
and the energy storage is in the form of potential energy 
of the metastable configurations of the grains,
which are possible thanks to the thresholds 
built by the static friction between grains.
When the input of grains makes one of the thresholds
to be overpassed, some grains start to move, 
this helps other grains to overpass their thresholds,
giving rise (or not) to an energy release in the form of an avalanche.

Earthquakes also fulfill this picture.
In this case the slow energy input comes from the relative motion 
of the tectonic plates, 
this energy is stored in the form of stress in the faults,
due again to the thresholds provided by static friction.
When an increase in stress cannot be sustained by friction, 
energy is released and redistributed in the system, 
triggering an avalanche of slips, i.e, an earthquake.

Among the geophysical phenomena mentioned at the end of Sec. 3
as displaying power-law statistics in their energy release,
most of them 
(rainfall, landslides, rock avalanches, forest fires, volcanic eruptions,
magnetosphere activity, and solar flares)
can be understood as self-organized critical systems,
see Table  \ref{tab:1}.
Perhaps, the only exceptions are tsunamis, 
which are not slowly driven
(but driven by earthquakes, landslides, etc.), 
and may be meteorites. 
In the next section we will discuss if the evolution of tropical cyclones
can be understood in these terms.

\begin{table}
\caption{
Self-organized critical characteristics of diverse phenomena.
All systems receive a slow driving of energy 
which is stored thanks to local thresholds;
eventually, a sudden release of energy spreads through the system.
}
\label{tab:1}       
%
%
\begin{center}
\begin{tabular}{l|cccc}
\hline\noalign{\smallskip}
           & sandpile          & earthquakes       & rainfall    & tropical cyclones \\
\noalign{\smallskip}\svhline\noalign{\smallskip}
driving    & addition          & motion of         & solar       & solar \\
           &  of grains        & tectonic plates   & radiation   & radiation\\
\noalign{\smallskip}\hline\noalign{\smallskip}
storage of \, & gravitational     & elastic           & water in    & heat of the sea \\
energy     & \, potential energy  \, & \, potential energy  \, & atmosphere  & \\
\noalign{\smallskip}\hline\noalign{\smallskip}
threshold  & friction          & friction          & saturation  & sea surface temperature$^a$  \\
\noalign{\smallskip}\hline\noalign{\smallskip}
spread of  & toppling of       & release           & nucleation  & wind \\
energy     & grains            & of stress         & of drops    & \\
\noalign{\smallskip}\svhline\noalign{\smallskip}
\end{tabular}
\\
\hfill $^a$ Plus a external trigger.
\end{center}
\end{table}

\section{Criticality of Tropical Cyclones}

Previously
we have shown that the energy dissipated by tropical cyclones
follows a power-law distribution.
As other catastrophic phenomena also show this behavior, 
and as some of these other complex phenomena can be accommodated to the 
perspective of self-organized criticality,
it is natural to investigate the possible connections
between self-organized criticality and tropical cyclones.

In principle, we can guarantee that 
the broad requirements of self-organized criticality 
are fulfilled in tropical cyclones.
Indeed, the tropical sea surface stores enormous quantities of energy,
in the form of warm water.
Naturally, this energy is slowly supplied to the sea by solar radiation.
Moreover, 
a certain amount of stored energy is necessary previous to its release
by a tropical cyclone, 
as if the sea surface temperature is below about $26^\circ$ C
these storms cannot develop \cite{Gray_storms}.
However, when the tropical cyclone is at work, 
the release of energy is very rapid 
(even more rapid compared with the slow heating of the sea by the sun).
This release is facilitated by
the strong winds, 
which increase the evaporation 
of water from the sea and then also the release of energy, 
which in turn increase the strength of the winds;
this is in some sense analogous to the chain-reaction nature 
of avalanches, in which part of the released energy is invested in 
facilitating further release.
It is important to stress that tropical cyclones 
liberate vast amounts of heat from the tropical oceans;
Emanuel estimates that quantity in
more than $10^{22}$ Joules every year 
\cite{Emanuel_bams08}.
Table \ref{tab:1} illustrates the energy flow of tropical 
cyclones comparing it with that of some well-known self-organized critical phenomena.

Nevertheless, there are also differences between tropical cyclones and
earthquakes or sandpiles.
In the latter cases, the release of energy spreads through the system
in all possible directions, in principle.
In contrast, a tropical cyclone attains a characteristic radius
and moves in an irregular but close to one-dimensional path, 
carried by the predominant large-scale winds.
Another difference is that favorable conditions, i.e., 
more than enough energy content in the sea, is not a sufficient condition
for these storms to develop.
As the experts know, some kind of perturbation is needed
to trigger the genesis process,
by means of easterly waves
for example \cite{Emanuel_book}.
So, some kind of overheating or supercriticality
seems to be present in the process.
Curiously, recent research seems to indicate
that most earthquakes do not occur ``spontaneously'' by the slow increase
of the tectonic stress, rather, they are triggered by the passing of seismic waves
\cite{Elst}.



\section{Tropical cyclone energy and climate change}

The mutual influence between global warming and tropical cyclones
constitutes a very complex issue.
In recent years, many works have investigated the response of tropical 
cyclones to increased sea surface temperature and other changing
climate indicators 
\cite{Emanuel_nature05,Goldenberg_Science,Trenberth,Landsea_comment,Webster_Science,Chan_comment,Klotzbach,Shepherd_Knutson,Kossin,Elsner08,Gray_comment,Landsea_Science06,Landsea_Eos07,Aberson,Elsner_Jagger}.
Most of these studies use measures that involve the change in the annual number of tropical cyclones,
as for instance the $PDI$ defined originally by Emanuel \cite{Emanuel_nature05}.
In contrast, the individual-cyclone $PDI$ probability distribution (introduced in the previous sections) 
is independent on the number of cyclones,
and allows the comparison of the characteristics 
of single events in different years \cite{Corral_hurricanes}.
This has the advantage of avoiding the count of the number of storms, 
which is severely underestimated in old records.

But what can one expect from the response of a self-organized critical system under a change 
in external conditions?
This kind of systems are supposed to show a robust behavior;
after all, the critical point is an attractor of the dynamics, 
which means that perturbations in the parameters that define the system are usually not relevant.
So, criticality, and therefore the power-law behavior, should hold
independently of the changing of climatic conditions.

That was indeed the result of Ref. \cite{Corral_hurricanes},
where it was shown that $D(PDI) \propto 1/PDI^\alpha$,
both for periods of high or low tropical-cyclone activity
or for periods of high or low sea surface temperature
(in the NAtl and EPac).
Does this mean that changing climate does not alter
the distribution of the energy released by tropical cyclones?
Not at all: although the power-law exponent $\alpha$ does not
change (under the statistical uncertainties)
the high$-PDI$ tail of the distribution does change.
Let us approximate the distribution by means of the following formula,
$$
D(PDI) \propto \frac{\exp(-PDI/a)}{PDI^\alpha},
$$
which covers both the power-law behavior for $PDI \ll a$ and the faster high$-PDI$ decrease,
modeled here by an exponential, for $PDI \gg a$.
The parameter $a$, called the cutoff, separates then
both behaviors.
A normalization constant, hidden under the proportionality symbol,
also depends on $a$, but this dependence is not important 
in our argument.

The effect of an increase in sea surface temperature
is just an increase in the value of $a$;
so, the transition from power-law behavior to
exponential decay occurs at a larger $PDI$ value
(given by $a$).
In other words, the $PDI$ values are shifted
by a scale factor equal to the ratio of increase of $a$.
As $a$ denotes the value of the $PDI$ for which tropical cyclones
are affected by the boundaries of the basin, we
can understand the increase in $a$ as an enlargement of
the effective size of the basin;
this is not surprising, higher sea surface temperature
implies that the part of the ocean over which tropical 
cyclones can develop is larger.

\begin{figure}[b]
\sidecaption
\includegraphics[scale=.55]{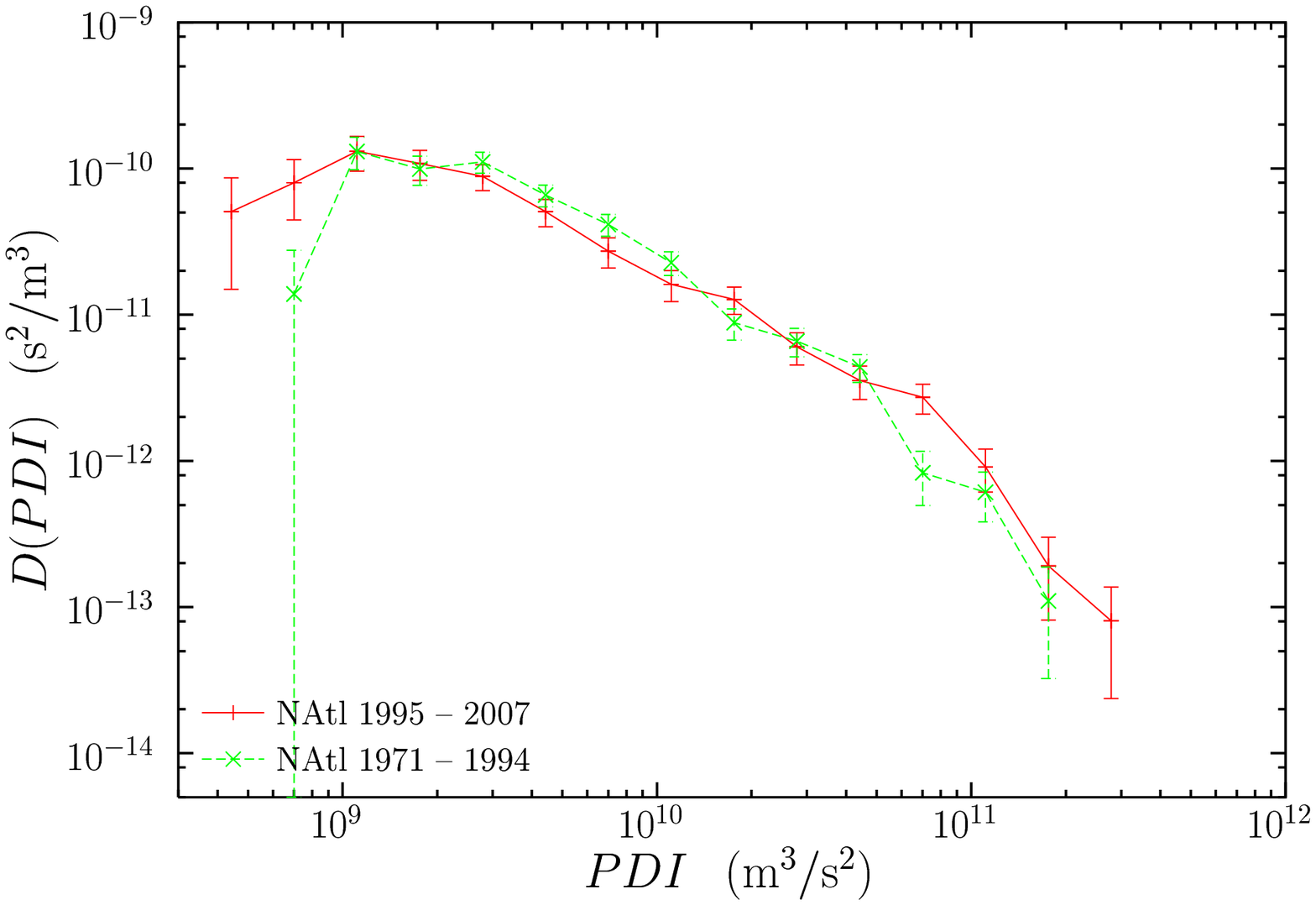}
\includegraphics[scale=.55]{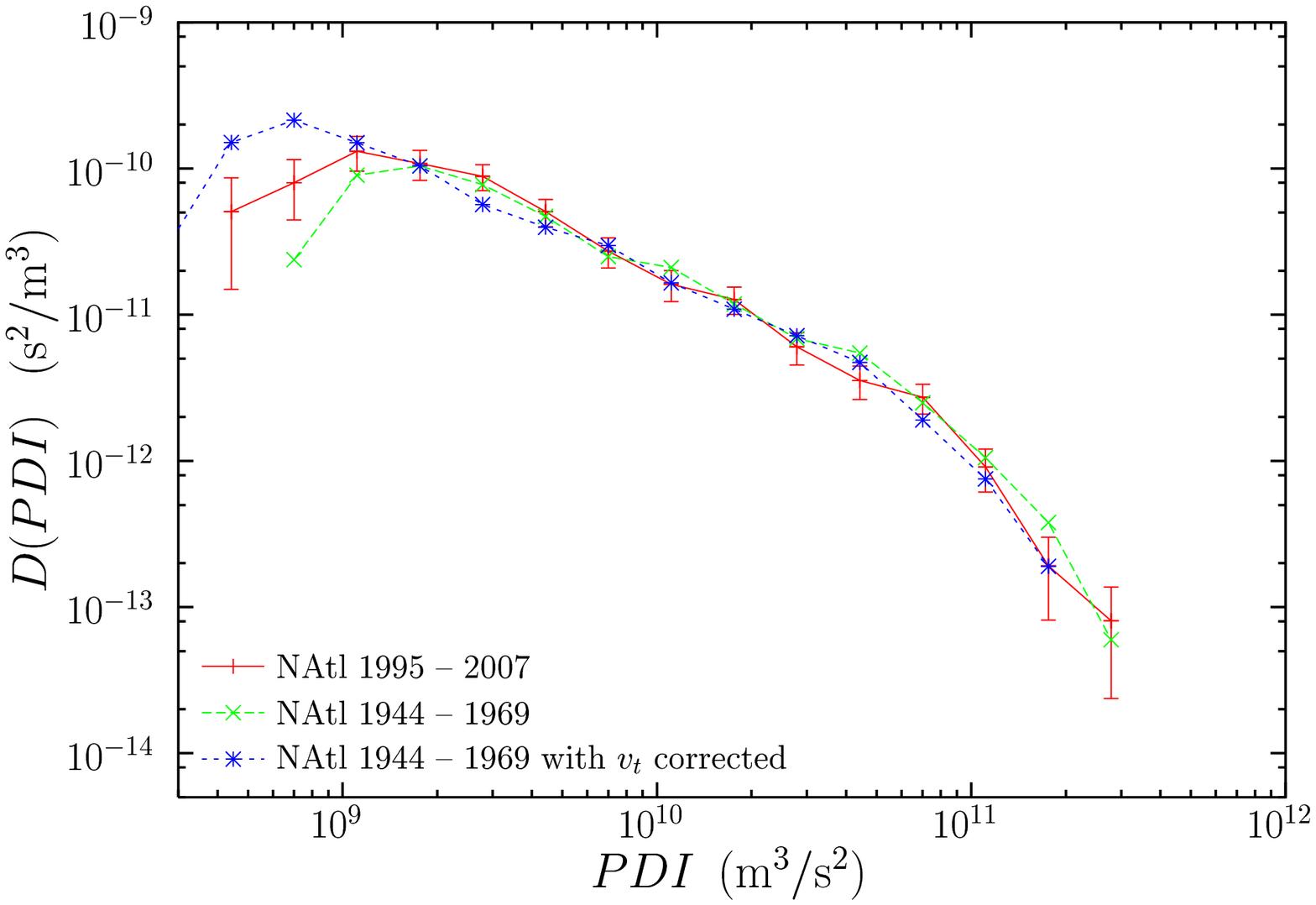}
\caption{
Probability density of tropical-cyclone $PDI$
for time periods with different levels of activity.
(a) The comparison between 1971-1994 and 1995-2007
shows that in the latter period larger $PDI$ values are possible.
(b) In contrast, the period 1944-1969 shows no significant
differences with 1995-2007. Even, a correction of the speeds of 
the former period (subtracting 4 m/s \cite{Landsea93}) does not change noticeably the results.
}
\label{fig:4}       
\end{figure}

Figure \ref{fig:4}(a) shows the $PDI$ distributions
in the North Atlantic
for the years 1971-1994 and 1995-2007.
The first period corresponds predominantly to relatively
low sea surface temperatures, whereas
in the second period the temperatures are higher.
We clearly see how the scale that delimits the boundary 
effects increases.
In summary, the last years of the North Atlantic are characterized
by larger hurricanes, in terms of dissipation of energy,
in comparison with the period 1971-1994.

Nevertheless, going back beyond 1970 yields a different tendency, 
as then the hurricanes show a distribution very similar to that of recent years.
In fact, Fig. \ref{fig:4}(b) compares the $PDI$ distribution for the period 1944-1969 
with the one corresponding to 1995-2007, showing no significant differences.
Even a correction of old values of the speeds inspired
in the work of Landsea \cite{Landsea93,Landsea_comment}, in which they are decreased
by an amount of 4 m/s, does not alter significantly the results.

\section{Discussion}

The criticality of tropical cyclones offers a new perspective 
for the understanding of this complex phenomenon.
Naturally, many questions arise, 
and much more research will be necessary to answer them.
First, we can wonder how this criticality relates to the 
results of Peters and Neelin \cite{Peters_np}, who have
recently proposed the criticality of the atmosphere for the
transition to rainfall occurrence 
(i.e., the transition from no precipitation to precipitation).
After all, tropical cyclones show, in addition to strong winds, 
enormous quantities of rainfall, and so they contribute to
the precipitation data analyzed by Peters and Neelin.
These authors showed that the state of the atmosphere, 
represented by its water-vapor content, 
is usually close to the onset of precipitation
(this onset marks the critical point of the transition).
However, tropical cyclones clearly surpass this onset of precipitation
(O. Peters, private communication) and then it is not clear why 
they still retain critical characteristics.

A subsequent question is how the idea of criticality 
affects our vision of atmospheric processes, 
and, in particular, the concept of a chaotic weather
\cite{Lorenz_book}.
It is true that both behaviors, chaos and criticality, 
share some characteristics, among them, 
an inherent unpredictability.
But there are also fundamental differences.
First, chaos usually appears in low-dimensional systems, 
i.e., systems described by a few non-linear differential 
equations, for instance, whereas criticality is the hallmark
of a high number of strongly interacting degrees of freedom.
And second, the unpredictability in chaos
is described by the exponential separation of close trajectories
(positive maximum Lyapunov exponent),
whereas in a critical system this divergence should be a power law
(with a zero maximum Lyapunov exponent, if one likes).
This suggest that a reconsideration of the limits of predictability
of the weather could give interesting outcomes
\cite{Orrell}.

\section{Acknowledgements}

The author was benefited by previous interaction with A. Oss\'{o},
made possible thanks to J. E. Llebot.
Feedback from
K. Emanuel,
E. Fukada,
J. Kossin, 
O. Peters, 
G. B. Raga, 
R. Romero, and 
A. Turiel
was very valuable.
Table 1 is based on a previous one
by O. Peters and K. Christensen.
Research projects 
2009SGR-164,
FIS2009-09508,
and specially FIS2007-29088-E,
from the EXPLORA - Ingenio 2010 program,
have contributed in one form or another to the execution
of the research.



\end{document}